\documentclass{my_elsart}
\usepackage{amsmath}

\newtheorem{theo}{Theorem}[section]
\newtheorem{coro}[theo]{Corollary}
\newtheorem{lemma}[theo]{Lemma}

\begin{document}
\begin{frontmatter}

\title{Some results on $(a:b)$-choosability}
\author[TelAviv]{Shai Gutner\corauthref{cor}},
\corauth[cor]{Corresponding author.}
\ead{gutner@tau.ac.il}
\author[TelAviv]{Michael Tarsi}
\ead{tarsi@post.tau.ac.il}
\address[TelAviv]{School of Computer Science, Tel-Aviv University, Tel-Aviv, 69978, Israel}

\begin{abstract}
A solution to a problem of Erd\H{o}s, Rubin and Taylor is obtained by showing
that if a graph $G$ is $(a:b)$-choosable, and $c/d > a/b$, then $G$ is not
necessarily $(c:d)$-choosable. 
Applying probabilistic methods, an upper bound
for the $k^{th}$ choice number of a graph is given. We also prove that a
directed graph with maximum outdegree $d$ and no odd directed
cycle is $(k(d+1):k)$-choosable for every $k \geq 1$.
Other results presented in this article are related to the strong choice number of graphs (a generalization of the strong chromatic number).
We conclude with complexity analysis of some decision problems related to graph choosability.
\end{abstract}

\end{frontmatter}

\section{Introduction}

The paper is based on the first author's master's thesis under the supervision of the second author \cite{Gu1}.
Many of the recent results on choosability can be found in the survey papers \cite{A3,T,KTV} and their many references.

All graphs considered are finite, undirected and simple.
A graph $G=(V,E)$ is {\em $(a:b)$-choosable} if for every family of 
sets $\{S(v): v \in V \}$, where $|S(v)|=a$ for all $v \in V$, there exist 
subsets $C(v) \subseteq S(v)$, where $|C(v)|=b$ for every $v \in V$, and 
$C(u) \cap C(v)= \emptyset$ whenever $u,v \in V$ are adjacent. 
The {\em $k^{th}$ choice number} of $G$, denoted by $ch_k(G)$,
is the smallest integer $n$ for which $G$ is $(n:k)$-choosable.
A graph $G=(V,E)$ is {\em $k$-choosable} if it is $(k:1)$-choosable.
The {\em choice number} of $G$, denoted by $ch(G)$, is equal to $ch_1(G)$.

The concept of $(a:b)$-choosability was defined and studied by Erd\H{o}s,
Rubin and Taylor in~\cite{ERT}. In the present paper we present some new results
related to that topic. Part of our work generalizes previous
results from ~\cite{A1},~\cite{A2}, ~\cite{AT} and ~\cite{ERT}. We list our results in this introduction section. The detailed proofs are given separately in later sections.

The following theorem examines the behavior of $ch_k(G)$ when
$k$ is large.
\begin{theo}
\label{t21}
Let G be a graph. For every $\epsilon >0$ there exists an integer $k_0$ such 
that for every $k \geq k_0$, $ch_k(G) \leq k(\chi (G)+\epsilon)$. 
\end{theo}
The following question is stated in~\cite{ERT}: \\
If $G$ is $(a:b)$-choosable, and $\frac{c}{d} > \frac{a}{b}$, does it 
imply that G is $(c:d)$-choosable? \\
The following is a negative answer to this question:
\begin{coro}
\label{c21}
If $l > m \geq 3$, then there exists a graph $G$ which is $(a:b)$-choosable
but not $(c:d)$-choosable,
where $\frac{c}{d} = l$ and $\frac{a}{b} = m$.
\end{coro}

Let $K_{m*r}$ denote the complete $r$-partite graph with $m$ vertices in
each vertex class, and let $K_{m_1, \ldots ,m_r}$ denote the complete
$r$-partite graph with $m_i$ vertices in the $i^{th}$ vertex class.
It is shown in~\cite{A2} that there exist two positive constants $c_1$
and $c_2$ such that 
$c_1 r \log m \leq ch(K_{m*r}) \leq c_2 r \log m$, for every $m \geq 2$ and $r \geq 2$. The following theorem
generalizes the upper bound.
\begin{theo}
\label{t31}
If $r \geq 1$ and $m_i \geq 2$ for every $1 \leq i \leq r$, then
\begin{equation*}
ch_k(K_{m_1, \ldots ,m_r}) \leq 948 r(k+\log{\frac{m_1+ \cdots + m_r}{r}}).
\end{equation*}
\end{theo}
Logarithms are in the natural base $e$.
Following are two applications of the above.
\begin{coro}
\label{c31}
For every graph G and $k \geq 1$
\begin{equation*}
ch_k(G) \leq 948  \chi (G) (k+\log{(\frac{|V|}{\chi(G)}+1)}).
\end{equation*}
\end{coro}
The second corollary generalizes a result from~\cite{A2} regarding the
choice numbers of random graphs. We refer to the standard model $G_{n,p}$
(see, e.g.,~\cite{Bo2}), a graph on $n$ vertices, every pair of which
is expected to be the endvertices of an edge,
randomly and independently, with probability $p$.  
\begin{coro}
\label{c32}
For every two constants $k \geq 1$ and $0 < p < 1$, the probability that 
$ch_k(G_{n,p}) \leq 475 \log{(1/(1-p))} n \frac{\log \log n}{\log n}$ tends 
to $1$ as $n$ tends to infinity.
\end{coro}

A theorem stated in~\cite{AT} reveals the connection between the
choice number of a graph and its orientations. We present here a
generalization of this theorem for a specific case:
\begin{theo}
\label{t41}
Let $D=(V,E)$ be a digraph and k a positive integer. For each $v \in V$, let $S(v)$
be a set of size $k(d^+_D(v)+1)$, where $d^+_D(v)$ is the outdegree of v.
If D contains no odd directed (simple) cycle, then there exist subsets 
$C(v) \subseteq S(v)$, where $|C(v)|=k$ for all $v \in V$, and 
$C(u) \cap C(v)= \emptyset$ for every two adjacent vertices $u,v \in V$.
The subsets $C(v)$ can be found in polynomial time with respect to $|V|$ and $k$.
\end{theo}
\begin{coro}
\label{c41}
Let G be an undirected graph. If $G$ has an orientation $D$ which contains 
no odd directed (simple) cycle and maximum outdegree d, then
$G$ is $(k(d+1):k)$-choosable for every $k \geq 1$.
\end{coro}
\begin{coro}
\label{c42}
An even cycle is $(2k:k)$-choosable for every $k \geq 1$.
\end{coro}
The last corollary enables us to generalize a variant of
Brooks' Theorem which appears in~\cite{ERT}.
\begin{coro}
\label{c43}
If a connected graph $G$ is not $K_n$, and not an odd cycle,
then $ch_k(G) \leq k \Delta(G)$ for every $k \geq 1$, where $\Delta(G)$
is the maximum degree of $G$.
\end{coro}
For a graph $G=(V,E)$, define $M(G)=\max(|E(H)|/|V(H)|)$,
where $H=(V(H),E(H))$ ranges over all subgraphs of $G$. The
following two corollaries are generalizations of results which appear
in~\cite{AT}.
\begin{coro}
\label{c44}
Every bipartite graph $G$ is
$(k(\lceil M(G)\rceil+1):k)$-choosable 
for every $k \geq 1$.
\end{coro}
\begin{coro}
\label{c45}
Every bipartite planar graph $G$ is $(3k:k)$-choosable for 
every $k \geq 1$.            
\end{coro}
Following are some more applications:
\begin{coro}
\label{c46}
If every induced subgraph of a graph $G$ has a vertex of degree at most $d$,
then $G$ is $(k(d+1):k)$-choosable for every $k \geq 1$.
\end{coro}
\begin{coro}
\label{c47}
If $G$ is a triangulated (chordal) graph, then $ch_k(G)=k\chi (G)=k\omega (G)$ for 
every $k \geq 1$, where $\omega (G)$ is the clique number of $G$.
\end{coro}

The list-chromatic conjecture asserts that for every graph $G$,
$ch(L(G))=\chi (L(G))$, where $L(G)$ denotes the line graph of $G$.
The list-chromatic conjecture is easy to verify for trees, graphs of       
degree at most 2, and $K_{2,m}$. It has also been proven true for
snarks~\cite{H}, $K_{3,3}$~\cite{CH}, $K_{4,4}$, $K_{6,6}$~\cite{AT}, 
and 2-connected regular planar graphs~\cite{EG}. 
Galvin proved the list-chromatic conjecture for all bipartite multigraphs \cite{Ga}.
The following corollary 
shows that the assertion of the list-chromatic conjecture is true 
for a graph whose 2-connected components are at most triangles:
\begin{coro}
\label{c48}
If a graph $G$ contains no simple circuit of size $4$ or more then
$ch(L(G))=\chi (L(G))$.
\end{coro}

The {\em core} of a graph $G$ is the graph obtained from $G$ by successively deleting
vertices of degree $1$ until there are no such vertices left.
The graph $\Theta_{a,b,c}$ consists of three paths of lengths $a$,$b$, and $c$, which share a pair of endvertices and they are otherwise vertex 
disjoint. 
The following theorem from~\cite{ERT} gives a complete characterization of 
$2$-choosable graphs:
\begin{theo}
\label{t51}
A connected graph $G$ is $2$-choosable if, and only if, the core of $G$ 
belongs to $\{ K_1 , C_{2m+2} , \Theta_{2,2,2m} : m \geq 1 \}$.
\end{theo} 
The following is asked in~\cite{ERT} : \\
If $G$ is $(a:b)$-choosable, does it follow that $G$ is $(am:bm)$-choosable? \\
The first instant of that question is proved in \cite{Gu1}, where it is shown that
if a graph $G$ is $2$-choosable, then $G$ is also $(4:2)$-choosable.
Tuza and Voigt later generalized this result by showing that
every $2$-choosable graph is $(2m:m)$-choosable \cite{TV}.

In the other direction we obtain:\\
\begin{theo}
\label{t53}
Suppose that $k$ and $m$ are positive integers and that $k$ is odd. If a graph 
$G$ is $(2mk:mk)$-choosable, then $G$ is also $2m$-choosable.
\end{theo}

A graph $G=(V,E)$ is {\em $f$-choosable} for a function $f: V \mapsto N$     
if for every family of sets $\{S(v): v \in V \}$, where $|S(v)|=f(v)$ for   
all $v \in V$, there is a proper vertex-coloring of $G$ assigning to
each vertex $v \in V$ a color from $S(v)$.
It is shown in~\cite{ERT} that the following problem is
$\Pi _2^p$-complete: ( for terminology see~\cite{GJ} )

\vspace{3 mm}
\noindent
{\bf BIPARTITE GRAPH $(2,3)$-CHOOSABILITY (BG $(2,3)$-CH)} \\
INSTANCE: A bipartite graph $G=(V,E)$ and a 
function $f: V \mapsto \{2,3\}$. \\
QUESTION: Is $G$ $f$-choosable? 

\vspace{3 mm}
\noindent
We consider the following decision problem:

\vspace{3 mm}
\noindent
{\bf BIPARTITE GRAPH $k$-CHOOSABILITY (BG $k$-CH)} \\
INSTANCE: A bipartite graph $G=(V,E)$. \\
QUESTION: Is $G$ $k$-choosable? 

\vspace{3 mm}
\noindent
If follows from theorem~\ref{t51} that this problem is solvable in
polynomial time for $k=2$. 
\begin{theo}
\label{t61}
{\bf BIPARTITE GRAPH $k$-CHOOSABILITY} is $\Pi _2^p$-complete for
every constant $k \geq 3$.
\end{theo}  
\noindent
Results concerning the complexity of planar graph choosability are proved in \cite{Gu2}.

A graph $G=(V,E)$ is {\em strongly $k$-colorable} if every graph obtained
from $G$ by appending a union of vertex disjoint cliques of size at     
most $k$ ( on the vertex set $V$ ) is $k$-colorable. 
An analogous definition of {\em strongly $k$-choosable} is made by replacing
colorability with choosability.
The {\em strong chromatic number} of a graph $G$, denoted by $s\chi(G)$, is
the minimum $k$ such that $G$ is strongly $k$-colorable. 
Define $s\chi(d) = \max(s\chi(G))$, where $G$ ranges over all graphs with
maximum degree at most $d$. 
(The definition of strong colorability given in~\cite{A1} is
slightly different. It is claimed there that if $G$ is strongly $k$-colorable,
then it is also strongly $(k+1)$-colorable. However, it is not known
how to prove this if the original definition given in~\cite{A1} is used).
\begin{theo}
\label{t71}
If $G$ is strongly $k$-colorable, then it is strongly $(k+1)$-colorable
as well.
\end{theo}
We give a weaker version of this theorem for choosability.
\begin{theo}
\label{t72}
If $G$ is strongly $k$-choosable, then it is also strongly $km$-choosable
for any integer $m$.
\end{theo}
\begin{theo}
\label{t73}
Let $G=(V,E)$ be a graph, and suppose that $km$ divides $|V|$. If the
choice number of any graph obtained from $G$ by appending a union of vertex 
disjoint $k$-cliques (on the vertex set $V$) is $k$, then the choice number
of any graph obtained from $G$ by appending a union of
vertex disjoint $km$-cliques is $km$.
\end{theo}
\begin{coro}
\label{c71}
Let $n$ and $k$ be positive integers, and let $G$ be a $(3k+1)$-regular graph
on $3kn$ vertices. Assume that $G$ has a decomposition into a Hamiltonian
circuit and $n$ pairwise vertex disjoint $3k$-cliques. Then $ch(G)=3k$.
\end{coro}

It is proved in~\cite{A1} that there is a constant $c$ such that for
every $d$, $3 \lfloor d/2 \rfloor < s\chi(d) \leq cd$. The following
theorem improves the lower bound.
\begin{theo}
\label{t74}
For every $d \geq 1$, $s\chi(d) \geq 2d$.
\end{theo}

\section{A solution to a problem of Erd\H{o}s, Rubin and Taylor}

In this section we prove an upper bound for the $k^{th}$ choice number of
a graph when $k$ is large and apply this bound to settle a problem
raised in~\cite{ERT}.

\noindent
{\bf Proof of Theorem~\ref{t21}}\,
Let $G=(V,E)$ be a graph and $\epsilon >0$. Let $r$ stand for the chromatic number of $G$ and let 
$\{ V_1,...,V_r \}$ be a partition of $V$ into stable sets.
Assign a set $S(v)$ of 
$\lfloor k(\chi (G)+\epsilon) \rfloor$ distinct colors to every $v \in V$. Let 
$S=\cup_{v \in V} S(v)$ be the set of all colors. Define
$R=\{1,2, \ldots ,r \}$ and let $f: S \mapsto R$ be a random function,
obtained by randomly selecting, 
the value of $f(c)$, independently for each color $c \in S$,   according to a uniform distribution on $R$. The colors
$c$ for which $f(c)=i$ will be used to color the 
vertices in $V_i$. To complete the proof, it thus suffices to
show that the probability of the following event is positive: For every $i$, $1 \leq i \leq r$, and
for every vertex $v \in V_i$ there are at least $k$ colors $c \in S(v)$
for which $f(c)=i$.

For a fixed vertex $v$, included in a set $V_i$, define 
$X=|S(v) \cap f^{-1}(i)|$.
The probability that there are less than $k$ colors $c \in S(v)$ for
which $f(c)=i$ is equal to $Pr(X<k)$. Since $X$ is a random variable with 
distribution $B(\lfloor k(r+\epsilon ) \rfloor  ,1/r)$, Chebyshev's 
inequality (see, e.g.,~\cite{AS}) implies
\begin{equation*}
Pr(X<k) \leq 
Pr(| X-\frac{\lfloor k(r+\epsilon) \rfloor }{r} | \geq
\frac{\lfloor k\epsilon \rfloor }{r} ) \leq
\frac{\lfloor k(r+\epsilon ) \rfloor \frac{1}{r} (1-\frac{1}{r})}
{(\frac{\lfloor k\epsilon \rfloor }{r})^2} =  O(\frac{1}{k}).
\end{equation*}
It follows that there is an integer $k_0$ such that for every 
$k \geq k_0$, $P(X<k)<1/|V|$. There are $|V|$ vertices from which $v$ is selected (and $i$ is determined)  
and hence, the probability that for some $i$ and some 
$v \in V_i$ there are less than $k$ colors $c \in S(v)$ for which $f(c)=i$
is smaller than $1$. $\Box$

\noindent
{\bf Proof of Corollary~\ref{c21}}\,
Suppose that $l > m \geq 3$, and let G be a graph such that $ch(G)=l+1$
and $\chi (G)=m-1$ ( it is proven in~\cite{V} that for every
$l \geq m \geq 2$ there is a graph $G$, where $ch(G)=l$ and $\chi (G)=m$.
Take for example the disjoint union of $K_m$ and $K_{n,n}$ for an appropriate
value of $n$ ). 
By theorem~\ref{t21}, for $\epsilon =1$ there exist an integer $k$ such
that $G$ is $(k(\chi(G)+1):k)$-choosable. Hence $G$ is
$(km:k)$-choosable but not $(l:1)$-choosable, as claimed. $\Box$

Note that it is not true that for every graph $G$ there exists an integer
$k_0$ such that $ch_k(G) \leq k \chi(G)$ for every $k \geq k_0$.
For example the chromatic number of $G=K_{3,3}$ is $2$, but that
graph is not $2$-choosable and therefore, by theorem~\ref{t53}, 
it is not $(2k:k)$-choosable for any odd $k$. Thus 
$ch_k(G) > k \chi(G)$ for every odd $k$. 

\section{An upper bound for the $k$th choice number}

In this section we establish an upper bound for 
$ch_k(K_{m_1, \ldots ,m_r})$, and use it to prove two consequences.
The following lemma appears in~\cite{AS}.
\begin{lemma}
\label{l31}
If $X$ is a random variable with distribution $B(n,p)$, $0 < p \leq 1$,
and $k<pn$ then
\begin{equation*}
Pr(X<k) < e^{-\frac{(np-k)^2}{2pn}}.
\end{equation*}
\end{lemma}

In the rest of this section we denote $t=\frac{m_1+ \cdots +m_r}{r}$,
$t_1=\frac{m_1+ \cdots +m_{r/2}}{r/2}$, and 
$t_2=\frac{m_{r/2+1}+ \cdots +m_r}{r/2}$.
Notice that $t=(t_1+t_2)/2$, and therefore $\log{t_1 t_2} \leq 2\log{t}$. 
\begin{lemma}
\label{l32}
If $1 \leq r \leq t$, $k \geq 1$, and $m_i \geq 2$ for every $i$,
$1 \leq i \leq r$, then    
$ch_k(K_{m_1, \ldots ,m_r}) \leq 4r(k+\log{t})$.
\end{lemma}
{\bf Proof}\,
Let $V_1, V_2, \ldots ,V_r$ be the stable sets of
$K = K_{m_1, \ldots ,m_r}$, where $|V_i|=m_i$ for all $i$, and let
$V = V_1 \cup \ldots \cup V_r$ be the set of all vertices of $K$.
For each $v \in V$, let $S(v)$ be a set of $\lfloor 4r(k+\log{t}) \rfloor$
distinct colors. Define $R=\{1,2, \ldots ,r \}$ and let $f: S \mapsto R$ be 
a random function, obtained by choosing, for each color $c \in S$, randomly
and independently, the value of $f(c)$ according to a uniform distribution  
on $R$. The colors $c$ for which $f(c)=i$ will be the ones to be used for   
coloring the vertices in $V_i$. To complete the proof it thus suffices to
show that with positive probability for every $i$, $1 \leq i \leq r$,
and every vertex $v \in V_i$ there are at least $k$ colors $c \in S(v)$
so that $f(c)=i$.

Fix an $i$ and a vertex $v \in V_i$, and define $X=|S(v) \cap f^{-1}(i)|$.
The probability that there are less than $k$ colors $c \in S(v)$ so
that $f(c)=i$ is equal to $Pr(X<k)$. Since $X$ is a random variable with
distribution $B(\lfloor 4r(k+\log{t}) \rfloor,1/r)$, by lemma~\ref{l31}
\begin{equation*}
Pr(X<k) < e^{-\frac{(E(X)-k)^2}{2E(X)}} \leq
e^{-\frac{(4(k+\log{t})-1-k)^2}{8(k+\log{t})}} <
\end{equation*}
\begin{equation*}
<
e^{-\frac{16(k+\log{t})^2-8(k+1)(k+\log{t})}{8(k+\log{t})}} \leq 
e^{-2\log{t}} = \frac{1}{t^2} \leq \frac{1}{rt},  
\end{equation*}
where the last inequality follows the fact that $r \leq t$. 
There are $rt$ possible ways to choose $i$, $1 \leq i \leq r$ and $v \in V_i$,
and hence, the probability that for some $i$ and some $v \in V_i$ there
are less than $k$ colors $c \in S(v)$ so that $f(c)=i$ is smaller than $1$,
completing the proof. $\Box$

\begin{lemma}
\label{l33}
Suppose that $r$ is even, $r > t$, $k \geq 1$, $d \geq 244$, and
$m_i \geq 2$ for every $i$, $1 \leq i \leq r$. If 
$ch_k(K_{m_1, \ldots ,m_{r/2}}) \leq 
d(1-\frac{1}{5 r^{1/3}})\frac{r}{2}(k+ \log{t_1})$ and  
$ch_k(K_{m_{r/2+1}, \ldots ,m_r}) \leq 
d(1-\frac{1}{5 r^{1/3}})\frac{r}{2}(k+ \log{t_2})$, then
$ch_k(K_{m_1, \ldots ,m_r}) \leq dr(k+ \log{t})$.
\end{lemma}
{\bf Proof}\,
Let $V_1, V_2, \ldots ,V_r$ be the stable sets of 
$K=K_{m_1, \ldots ,m_r}$, where $|V_i|=m_i$ for all $i$, and let
$V=V_1 \cup \ldots \cup V_r$ be the vertex set of $K$.
For each $v \in V$, let $S(v)$ be a set of $\lfloor dr(k+ \log{t}) \rfloor$
distinct colors. Define $R=\{1,2, \ldots ,r\}$, and let 
$S=\cup_{v \in V} S(v)$ be the set of all colors. Define 
$R_1=\{1,2, \dots ,r/2\}$ and $R_2=\{r/2+1, \ldots ,r\}$.    
Let $f:S \mapsto \{1,2\}$ be a random function obtained by choosing, for
each $c \in S$ randomly and independently, $f(c) \in \{1,2\}$ where for
all $j \in \{1,2\}$ 
\begin{equation*}
Pr(f(c)=j)=\frac{k+\log{t_j}}{2k+\log t_1 t_2}.
\end{equation*}
The colors $c$ for which $f(c)=1$ will be used for coloring the vertices
in $\cup_{i \in R_1} V_i$, whereas the colors $c$ for which $f(c)=2$
will be used for coloring the vertices in $\cup_{i \in R_2} V_i$.

For every vertex $v \in V$, define $C(v)=S(v) \cap f^{-1}(1)$ if $v$
belongs to $\cup_{i \in R_1} V_i$, and $C(v)=S(v) \cap f^{-1}(2)$
if $v$ belongs to $\cup_{i \in R_2} V_i$. Because of the assumptions of 
the lemma, it remains to show that with positive probability,
\begin{equation}
\label{e31}
|C(v)| \geq d(1-\frac{1}{5 r^{1/3}})\frac{r}{2}(k+\log{t_j})
\end{equation}
for all $j \in \{1,2\}$ and $v \in \cup_{i \in R_j} V_i$.

Fix a $j \in \{1,2\}$ and a vertex $v \in \cup_{i \in R_j} V_i$, and 
define $X=|C(v)|$. The expectation of $X$ is
\begin{equation*}
\lfloor dr(k+\log{t}) \rfloor \frac{k+\log{t_j}}{2k+\log t_1 t_2} \geq
(dr(k+\log{t}) -1) \frac{k+\log{t_j}}{2k+2\log{t}} \geq
d\frac{r}{2}(k+\log{t_j}) -1 = T.
\end{equation*}
If follows from lemma~\ref{l31} and the inequality $E(X) \geq T$ that
\begin{equation*}
Pr(X<T-T^{2/3})<e^{-\frac{(E(X)-T+T^{2/3})^2}{2E(X)}} \leq
e^{-\frac{1}{2} T^{1/3}} \leq e^{-\frac{1}{2}(d\frac{r}{2})^{1/3}}.
\end{equation*}
Since $|\cup_{i \in R_j} V_i| \leq rt < r^2$, the probability that 
$|C(v)|<T-T^{2/3}$ holds for some $v \in \cup_{i \in R_j} V_i$ 
is at most
\begin{equation*}
r^2 \cdot e^{-\frac{1}{2}(d\frac{r}{2})^{1/3}} < 1/2 ,
\end{equation*}
where the last inequality follows the fact that $d \geq 244$.
One can easily verify that
\begin{equation*}
T-T^{2/3}=T(1-\frac{1}{T^{1/3}}) \geq 
d\frac{r}{2}(k+\log{t_j})(1-\frac{1}{5 r^{1/3}}),
\end{equation*}
and therefore, with positive probability (\ref{e31}) holds for all 
$j \in \{1,2\}$ and $v \in \cup_{i \in R_j} V_i$. $\Box$

\noindent
{\bf Proof of Theorem~\ref{t31}}\,
Define for every $r$ which is a power of $2$	
\begin{equation*}
f(r)=\prod_{j=0}^{\log_2 r} (1-\frac{1}{5 \cdot 2^{j/3}}) /
\prod_{j=0}^{2} (1-\frac{1}{5 \cdot 2^{j/3}}).
\end{equation*}
We claim that for every $r$ which is a power of $2$
\begin{equation}
\label{e32}
ch_k(K_{m_1, \ldots ,m_r}) \leq \frac{244r(k+\log{t})}{f(r)}.
\end{equation}
The proof is by induction on $r$.

\noindent
{\bf Case 1:}\, $r \leq t$.

\noindent
The result follows from lemma~\ref{l32} since
\begin{equation*}
\frac{244}{f(r)} \geq
244 \prod_{j=1}^{2} (1-\frac{1}{5 \cdot 2^{j/3}}) > 4.
\end{equation*}

\noindent
{\bf Case 2:}\, $r > t$.

\noindent 
Notice that $t \geq 2$, and therefore $r \geq 4$. By the induction hypothesis 
\begin{equation*}
ch_k(K_{m_1, \ldots ,m_{r/2}}) \leq 
\frac{244(1-\frac{1}{5 r^{1/3}})\frac{r}{2}(k+\log{t_1})}{f(r)}
\end{equation*}
and
\begin{equation*}
ch_k(K_{m_{r/2+1}, \ldots ,m_r}) \leq
\frac{244(1-\frac{1}{5 r^{1/3}})\frac{r}{2}(k+\log{t_2})}{f(r)}.
\end{equation*}
Since $r \geq 4$, we have $244/f(r) \geq 244$ and it follows from lemma
\ref{l33} that (\ref{e32}) holds, as claimed.

It is easy to check that
\begin{equation*}
\prod_{j=3}^{\log_2 r} (1-\frac{1}{5 \cdot 2^{j/3}}) \geq
1 - \sum_{j=3}^{\log_2 r} \frac{1}{5 \cdot 2^{j/3}} \geq
1 - \frac{1}{10(1 - 2^{-1/3})},    
\end{equation*}
and therefore $244/f(r) \leq 474$. It follows from (\ref{e32}) that
for every $r$ which is a power of $2$
\begin{equation}
\label{e33}
ch_k(K_{m_1, \ldots ,m_r}) \leq 474r(k+\log{t}).
\end{equation}

Returning to the general case, assume that $r \geq 1$. Choose an integer
$r'$ which is a power of $2$ and $r \leq r' < 2r$. By applying
(\ref{e33}),  we get
\begin{equation*}
ch_k(K_{m_1, \ldots ,m_r}) \leq
ch_k(K_{m_1, \ldots ,m_r, \underbrace{2, \ldots ,2}_{r'-r}})
\end{equation*}
\begin{equation*}
\leq 474 r'(k+\log{\frac{m_1 + \cdots + m_r + 2(r'-r)}{r'}}) \leq
948 r(k+\log{\frac{m_1 + \cdots + m_r}{r}}),
\end{equation*}
completing the proof. $\Box$

Define $K = K_{m, \underbrace{s, \ldots ,s}_{r}}$, where $m \geq 2$ and
$s \geq 2$. Every induced subgraph of $K$ has a vertex of degree at
most $rs$, and therefore by corollary~\ref{c46}
$ch_k(K) \leq k(rs+1)$ for all $k \geq 1$. Note that this upper bound 
for $ch_k(K)$ does not depend of $m$, which means that a good lower bound 
for $ch_k(K_{m_1, \ldots ,m_r})$ has a more complicated form than the 
upper bound given in theorem~\ref{t31}.

\noindent
{\bf Proof of Corollary~\ref{c31}}\,
Let $G=(V,E)$ be a graph and $k \geq 1$. Define $r=\chi (G)$, and let
$V=V_1 \cup \ldots \cup V_r$ be a partition of the vertices, such that
each $V_i$ is a stable set. Define $m_i=|V_i|$ for all 
$i$, $1 \leq i \leq r$. By theorem~\ref{t21}
\begin{equation*}
ch_k(G) \leq ch_k(K_{m_1+1, \ldots ,m_r+1}) \leq
948r(k+\log{\frac{m_1+ \cdots + m_r + r}{r}}) =
\end{equation*}
\begin{equation*}
=
948\chi (G) (k+\log{(\frac{|V|}{\chi(G)}+1)}),
\end{equation*}
as claimed. $\Box$

\noindent
{\bf Proof of Corollary~\ref{c32}}\,
As proven by Bollob\'as in~\cite{Bo1}, for a fixed probability $p$,
$0 < p < 1$, almost surely (i.e., with probability that tends to $1$ as 
$n$ tends to infinity), the random graph $G_{n,p}$ has chromatic number 
\begin{equation*}
(\frac{1}{2}+o(1)) \log{(1/(1-p))}\frac{n}{\log n }.
\end{equation*}
By corollary~\ref{c31}, for every $\epsilon > 0$ almost surely
\begin{equation*}
ch_k(G_{n,p}) \leq 948(\frac{1}{2}+\epsilon)\log{(1/(1-p))}\frac{n}{\log n}
(k+\log{(\frac{3\log n}{\log{(1/(1-p))}}+1)}).
\end{equation*}
The result follows since $k$ and $p$ are constants. $\Box$

\noindent
Note that in the proof of the last corollary we have not used any knowledge 
concerning independent sets of $G_{n,p}$, as was done in~\cite{A2} for the
proof of the special case.

\section{Choice numbers and orientations}

Let $D=(V,E)$ be a digraph. We denote the set of out-neighbors of $v$ 
in $D$ by $N^+_D(v)$. A set of vertices $K \subseteq V$ is called a
{\em kernel} of $D$ if $K$ is an independent set and  
$N^+_D(v) \cap K \neq \emptyset$ for every vertex $v \not\in K$. 
Richardson's theorem (see, e.g.,~\cite{Be}) states that any digraph with
no odd directed cycle has a kernel.

\noindent
{\bf Proof of Theorem~\ref{t41}}\,
Let $D=(V,E)$ be a digraph which contains no odd directed (simple) cycle 
and $k \geq 1$. For each $v \in V$, let $S(v)$ be a set of size
$k(d^+_D(v)+1)$. We claim that the following algorithm finds subsets 
$C(v) \subseteq S(v)$, where $|C(v)|=k$ for all $v \in V$, and
$C(u) \cap C(v) = \emptyset$ for every two adjacent vertices $u,v \in V$.
\begin{enumerate}
\item
\label{s1}
$S \leftarrow \cup_{v \in V} S(v)$, $W \leftarrow V$ and for
every $v \in V$, $C(v) \leftarrow \emptyset$.
\item
\label{s2}
Choose a color $c \in S \cap \cup_{v \in W} S(v)$ and put
$S \leftarrow S - \{c\}$. 
\item
\label{s3}
Let $K$ be a kernel of the induced subgraph of $D$ on the vertex set
$\{ v \in W : c \in  S(v) \}$.
\item
\label{s4}
$C(v) \leftarrow C(v) \cup \{c\}$ for all $v \in K$.
\item
\label{s5}
$W \leftarrow W - \{ v \in K : |C(v)|=k \}$.
\item
\label{s6}
If $W = \emptyset$, stop. If not, go to step~\ref{s2}. 
\end{enumerate}

During the algorithm, $W$ is equal to $\{v \in V : |C(v)|<k \}$, and $S$ is
the set of remaining colors. We first prove that in step~\ref{s2},
$S \cap \cup_{v \in W} S(v) \neq \emptyset$. When the algorithm reaches
step~\ref{s2}, it is obvious that $W \neq \emptyset$. Suppose that $w \in W$ in 
this step, and therefore $|C(w)|<k$. It follows easily from the definition of
a kernel that every color from $S(w)$, which has been previously chosen in
step~\ref{s2}, belongs either to $C(w)$ or to $\cup_{v \in N^+_D(w)} C(v)$.
Since 
\begin{equation*}
|C(w)| + |\bigcup_{v \in N^+_D(w)} C(v)| < k + k \cdot d^+_D(w) = |S(w)|,
\end{equation*}
not all the colors of $S(w)$ have been used. This means that 
$S \cap S(w) \neq \emptyset$, as needed. It follows easily that the
algorithm always terminates.

Upon termination of the algorithm, $|C(v)|=k$ for all $v \in V$. In 
step~\ref{s4} the same color is assigned to the vertices of a kernel which is 
an independent set, and therefore $C(u) \cap C(v) = \emptyset$ for every two 
adjacent vertices $u,v \in V$. This proves the correctness of the algorithm.

In step~\ref{s4}, the operation $C(v) \leftarrow C(v) \cup \{c\}$ is performed 
for at least one vertex. Upon termination $|\cup_{v \in V} C(v)| \leq k|V|$, 
which means that the algorithm performs at most $k|V|$ iterations. There is
a polynomial time algorithm for finding a kernel in a digraph with no odd 
directed cycle. Thus, the algorithm is of polynomial time complexity 
in $|V|$ and $k$, completing the proof. $\Box$

\noindent
{\bf Proof of Corollary~\ref{c41}}\,
This is an immediate consequence of theorem~\ref{t41}, since
$k(d^+_D(v)+1) \leq k(d+1)$ for every $v \in V$. $\Box$

\noindent
{\bf Proof of Corollary~\ref{c42}}\,
The result follows from~\ref{c41} by taking the cyclic orientation of
the even cycle. $\Box$

The proof of corollary~\ref{c43} is similar to the proof of the special
case which appears in~\cite{ERT}. A graph $G=(V,E)$ is
{\em $k$-degree-choosable} if for every family of sets $\{S(v): v \in V \}$,
where $|S(v)|=k d(v)$ for all $v \in V$, there are subsets 
$C(v) \subseteq S(v)$, where $|C(v)|=k$ for all $v \in V$, and 
$C(u) \cap C(v)= \emptyset$ for every two adjacent vertices $u,v \in V$.  
\begin{lemma}
\label{l41}
If a graph $G=(V,E)$ is connected, and $G$ has a connected induced subgraph 
$H=(V',E')$ which is $k$-degree-choosable, then $G$ is $k$-degree-choosable.
\end{lemma}
{\bf Proof}\,
For each $v \in V$, let $S(v)$ be a set of size $k d(v)$. The proof is 
by induction on $|V|$. In case $|V|=|V'|$ there is nothing to prove. Assuming
that $|V|>|V'|$, let $v$ be a vertex of $G$ which is at maximal distance   
from $H$. This guarantees that $G - v$ is connected. Choose any subset
$C(v) \subseteq S(v)$ such that $|C(v)|=k$, and remove the colors of
$C(v)$ from all the vertices adjacent to $v$. The choice can be completed
by applying the induction hypothesis on $G - v$. $\Box$
\begin{lemma}
\label{l42}
If $c \geq 2$, then $\Theta_{a,b,c}$ is $k$-degree-choosable for every
$k \geq 1$.
\end{lemma}
{\bf Proof}\,
Suppose that $\Theta_{a,b,c}$ has vertex set 
$V=\{u,v,x_1,\ldots,x_{a-1},y_1,\ldots,y_{b-1},\\ z_1,\ldots,z_{c-1}\}$ and
contains the three paths 
$u-x_1-\cdots-x_{a-1}-v$, $u-y_1-\cdots-y_{b-1}-v$, and
$u-z_1-\cdots-z_{c-1}-v$. 
For each $w \in V$, let $S(w)$ be a set of size $kd(w)$. For the vertex
$u$ we choose a subset $C(u) \subseteq S(u) - S(z_1)$ of size $k$. 
For each  vertex according to the sequence 
$x_1,\ldots,x_{a-1},y_1,\ldots,y_{b-1},v,z_{c-1},\ldots,z_1$ we choose  
a subset of $k$ colors that were not chosen in adjacent earlier  vertices.
$\Box$

\noindent
For the proof of corollary~\ref{c43}, we shall need the following lemma
which appears in~\cite{ERT}.
\begin{lemma}
\label{l43}
If there is no  vertex which disconnects $G$, then $G$ is an odd cycle, or   
$G=K_n$, or $G$ contains, as a  vertex induced subgraph, an even cycle without
chord or with only one chord.
\end{lemma}     

\noindent
{\bf Proof of Corollary~\ref{c43}}\,
Suppose that a connected graph $G$ is not $K_n$, and not an odd cycle.
If $G$ is not a regular graph, then every induced subgraph of $G$ has 
a vertex of degree at most $\Delta(G)-1$, and by corollary~\ref{c46}
$ch_k(G) \leq k \Delta(G)$ for all $k \geq 1$. 
If $G$ is a regular graph, then there is  
a part of $G$ not disconnected by a  vertex, which is neither an odd cycle
nor a complete graph. It follows from lemma~\ref{l43} that $G$ contains,
as a  vertex induced subgraph, an even cycle or a particular kind of 
$\Theta_{a,b,c}$ graph. We know from corollary~\ref{c42} and lemma~\ref{l42}
that both an even cycle and $\Theta_{a,b,c}$ are $k$-degree-choosable for
every $k \geq 1$. The result follows from lemma~\ref{l41}. $\Box$

\noindent
{\bf Proof of Corollary~\ref{c44}}\,
It is proved in~\cite{AT} that a graph $G=(V,E)$ has an orientation $D$ in
which every outdegree is at most $d$ if and only if $M(G) \leq d$. Therefore,
there is an orientation $D$ of $G$ in which the maximum outdegree is at
most $\lceil M(G) \rceil$. Since $D$ contains no odd directed cycles, the
result follows from corollary~\ref{c41}. $\Box$

\noindent
{\bf Proof of Corollary~\ref{c45}}\,
$M(G) \leq 2$, since any planar bipartite (simple) graph 
on $r$ vertices contains
at most $2r-2$ edges. The result follows from corollary~\ref{c44}. $\Box$

\noindent
{\bf Proof of Corollary~\ref{c46}}\,
We claim that if every induced subgraph of a graph $G=(V,E)$ has a vertex of
degree at most $d$, then $G$ has an acyclic orientation in which the 
maximum outdegree is $d$. The proof is by induction on $|V|$. If $|V|=1$,
the result is trivial. If $|V|>1$, let $v$ be a vertex of $G$ with degree
at most $d$. By the induction hypothesis, $G - v$ has an acyclic orientation
in which the maximum outdegree is $d$. We complete this orientation of 
$G - v$ by orienting every edge incident to $v$ from $v$ to its appropriate
neighbor and obtain the desired orientation of $G$, as claimed. The result 
follows from corollary~\ref{c41}. $\Box$

An undirected graph $G$ is called {\em triangulated} if $G$ does not
contain an induced subgraph isomorphic to $C_n$ for $n \geq 4$. Being 
triangulated is a hereditary property inherited by all the induced 
subgraphs of $G$. A vertex $v$ of $G$ is called {\em simplicial} if its
adjacency set $Adj(v)$ induces a complete subgraph of $G$. It is proved 
in~\cite{G} that every triangulated graph has a simplicial vertex.

\noindent
{\bf Proof of Corollary~\ref{c47}}\,
Suppose that $G$ is a triangulated graph, and let $H$ be an induced subgraph
of $G$. Since $H$ is triangulated, it has a simplicial vertex $v$.
The set of vertices $\{v\} \cup Adj_H(v)$ induces a complete subgraph of $H$,
and therefore $v$ has degree at most $\omega(G)-1$ in $H$. It follows
from corollary~\ref{c46} that $ch_k(G) \leq k\omega(G)$ for
every $k \geq 1$.
For every graph $G$ and $k \geq 1$, $ch_k(G) \geq k\omega(G)$ and hence
$ch_k(G)=k\omega(G)$ for every $k \geq 1$. Since $G$ is triangulated, it
is also perfect, which means that $\chi(G)=\omega(G)$, as needed. $\Box$

\noindent
{\bf Proof of Corollary~\ref{c48}}\,
It is easy to see that $L(G)$ is triangulated if and only if $G$ contains
no $C_n$ for every $n \geq 4$. The result follows from
corollary~\ref{c47}. $\Box$

The validity of the list-chromatic conjecture for graphs of class $2$ with
maximum degree $3$ (and in particular for snarks) follows easily from
corollary~\ref{c43}. Suppose that $G$ is a graph of class $2$ with 
$\Delta(G)=3$. Let $C$ be a connected component of $L(G)$. If $C$ is not 
a complete graph, and not an odd cycle, then 
$ch(C) \leq \Delta(C) \leq \Delta(L(G)) \leq 4$.
If $C$ is a complete graph or an odd cycle, then it is easy to see
that $\Delta(C) \leq 2$, and therefore by corollary~\ref{c46}
$ch(C) \leq \Delta(C)+1 \leq 3$. It follows that $ch(L(G)) \leq 4$.
Since $G$ is a graph of class $2$,
$ch(L(G)) \geq \chi(L(G)) = \Delta(G)+1 = 4$, and hence,
$ch(L(G))= \chi(L(G)) = 4$.

\section{Properties of $(2k:k)$-choosable graphs}

In this section we establish an upper bound for the
choice number of $(2k:k)$-choosable graphs.

\noindent
{\bf Proof of Theorem~\ref{t53}}\,
Suppose that $G=(V,E)$ is $(2mk:mk)$-choosable for $k$ odd. We prove that 
$G$ is $2m$-choosable as well. For each $v \in V$, let $S(v)$ be a set of
size $2m$. With every color $c$ we associate a set $F(c)$ of size $k$, such
that $F(c) \cap F(d) = \emptyset$ if $c \neq d$. For every $v \in V$,
we define $T(v)=\cup_{c \in S(v)} F(c)$. Since $G$ is $(2mk:mk)$-choosable, 
there are subsets $C(v) \subseteq T(v)$, where $|C(v)|=mk$ for 
all $v \in V$, and $C(u) \cap C(v) = \emptyset$ for every
two adjacent vertices $u,v \in V$.

Fix a vertex $v \in V$. Since $k$ is odd, there is a color $c \in S(v)$
for which $|C(v) \cap F(c)| > k/2$, so we define $f(v)=c$. 
In case $u$ and $v$ are adjacent vertices for which $c \in S(u) \cap S(v)$,
it is not possible that both $|C(u) \cap F(c)|$ and $|C(v) \cap F(c)|$ are
greater than $k/2$. This proves that $f$ is a proper vertex-coloring of
$G$ assigning to each vertex $v \in V$ a color in $S(v)$. $\Box$

\section{The complexity of graph choosability}

Let $G=(V,E)$ be a graph. We denote by $G'$ the graph obtained from $G$ by
adding a new vertex to $G$, and joining it to every vertex in $V$.  
Consider the following decision problem:

\vspace{3 mm}
\noindent
{\bf GRAPH $k$-COLORABILITY} \\
INSTANCE: A graph $G=(V,E)$. \\
QUESTION: Is $G$ $k$-colorable?

\vspace{3 mm}
\noindent
The standard technique to show a polynomial transformation from
GRAPH $k$-COLORABILITY to GRAPH $(k+1)$-COLORABILITY is to use the fact 
that $\chi(G') = \chi(G)+1$ for every graph $G$. 
However, it is not true that $ch(G') = ch(G)+1$ for every graph $G$.
To see that, we first prove that $K_{2,4}'$ is $3$-choosable.  

Suppose that $K_{2,4}'$ has vertex set $V=\{v,x_1,x_2,y_1,y_2,y_3,y_4\}$,
and contains exactly the edges $\{x_i,y_j\}$, $\{v,x_i\}$, and $\{v,y_j\}$.
For each $w \in V$, let $S(w)$ be a set of size $3$. 

\noindent
{\bf Case 1:}\, All the sets are the same.

\noindent
A choice can be made since $K_{2,4}'$ is 3-colorable.

\noindent
{\bf Case 2:}\, There is a set $S(x_i)$ which is not equal to $S(v)$.

\noindent
Without loss of generality, suppose that $S(v) \neq S(x_1)$.
For the  vertex $v$, choose a color $c \in S(v)-S(x_1)$, and remove $c$ from
the sets of the other vertices. We can assume that every set $S(y_j)$
is of size $2$ now.

Suppose first that $S(x_1)$ and $S(x_2)$ are disjoint. The number of
different sets consisting of one color from each of the $S(x_i)$ is
at least 6, and therefore we can choose colors $c_i \in S(x_i)$, such 
that $\{c_1,c_2\}$ does not appear as a set of $S(y_j)$.  
We complete the choice by choosing for every vertex $y_j$ a color 
from $S(y_j)-\{c_1,c_2\}$. 
Suppose next that $d \in S(x_1) \cap S(x_2)$. For every vertex $x_i$ we 
choose $d$, and for every vertex $y_j$ we choose a color from $S(y_j)-\{d\}$.

\noindent
{\bf Case 3:}\, There is a set $S(y_j)$ which is not equal to $S(v)$.

\noindent
Without loss of generality, suppose that $S(v) \neq S(y_1)$.
For the  vertex $v$, choose a color $c \in S(v)-S(y_1)$, and remove $c$ from 
the sets of the other vertices. 
Suppose first that $S(x_1)$ and $S(x_2)$ are disjoint. The number of
different sets consisting of one color from each of the $S(x_i)$ is 
at least 4, and since $|S(y_1)|=3$ we can choose colors $c_i \in S(x_i)$, 
such that $S(y_j)-\{c_1,c_2\} \neq \emptyset$ for every vertex $y_j$.
We can complete the choice as in case 2.
In case $S(x_1)$ and $S(x_2)$ are not disjoint, we proceed as in case 2. 

This completes the proof that $K_{2,4}'$ is 3-choosable. It follows from
theorem~\ref{t51} and corollary~\ref{c46} that $ch(K_{2,4})=3$, and
therefore $ch(K_{2,4}')=ch(K_{2,4})=3$. The following lemma exhibits
a construction which increases the choice number of a graph by exactly $1$.
\begin{lemma}
\label{l61}
Let $G=(V,E)$ be a graph. If $H$ is the disjoint union 
of $|V|+1$ copies of $G$,
then $ch(H')=ch(G)+1$.
\end{lemma}
{\bf Proof}\,
Let $H$ be the disjoint union of the graphs 
$\{G_i : 1 \leq i \leq |V|+1\}$, 
where each $G_i$ is a copy of $G$. 
Suppose that $H'$ is obtained from $H$ by joining the new vertex $v$ to all
the vertices of $H$.

We claim that if $G$ is $k$-choosable, then $H'$ is $(k+1)$-choosable.
For each $w \in V(H')$, let $S(w)$ be a set of size $k+1$. Choose a color
$c \in S(v)$, and remove $c$ from the sets of the other vertices. We can
complete the choice since $G$ is $k$-choosable.

We now prove that if $H'$ is $k$-choosable, then 
$G$ is $(k-1)$-choosable. 
By above, $ch(H') \leq ch(G)+1 \leq |V|+1$.   
Hence, we can assume that $k \leq |V|+1$.
For each $w \in V$, let $S(w)$ be a set of size $k-1$.
Without loss of generality $S(w) \cap \{1,2, \ldots , |V|+1\} = \emptyset$. 
For every $i$, $1 \leq i \leq |V|+1$, on the vertices of the graph $G_i$ we 
put the sets $S(w)$ together with the additional color $i$. The 
vertex $v$ is given the set $\{1,2, \ldots ,k\}$. 
Let $f$ be a proper vertex-coloring of $H'$ assigning to each vertex 
a color from its set. Denote $f(v)=i$, then $f$ restricted to $G_i$ is
a proper vertex-coloring of~$G$ assigning to each vertex $w \in V$
a color in S(w). 
By corollary~\ref{c43}, $ch(G)+1 \leq |V|$ if $G$ is not a complete
graph and the proof still goes through if $|V|+1$ is replaced by $|V|$
in the statement of the lemma. $\Box$  
\begin{lemma}
\label{l62}
{\bf BIPARTITE GRAPH $3$-CHOOSABILITY} is $\Pi_2^p$-complete.
\end{lemma}
{\bf Proof}\,
It is easy to see that {\bf BG $3$-CH} $\in \Pi_2^p$.
We transform {\bf BG $(2,3)$-CH} to {\bf BG $3$-CH}. 
Let $G=(V,E)$ and $f: V \mapsto \{2,3\}$ be an
instance of {\bf BG $(2,3)$-CH}.  
We shall construct a bipartite graph $H''$ such that $H''$ is $3$-choosable
if and only if $G$ is $f$-choosable.

Let $H$ be the disjoint union of the graphs $\{G_{i,j} : 1 \leq i,j \leq 3\}$,
where each $G_{i,j}$  is a copy of $G$. Let $(X,Y)$ be a bipartition of the
bipartite graph $H$. The graph $H''$ is obtained from $H$ by adding two new
vertices $u$ and $v$, joining $u$ to every vertex $w \in X$ for which 
$f(w)=2$, and joining $v$ to every vertex $w \in Y$ for which $f(w)=2$.

Since $H$ is bipartite, $H''$ is also a bipartite graph. It is easy to see that 
if $G$ is $f$-choosable, then $H''$ is $3$-choosable. We now prove that
if $H''$ is $3$-choosable, then $G$ is $f$-choosable. 
For every $w \in V$, let $S(w)$ be a set of size $f(w)$, such that 
$S(w) \cap \{1,2,3\} = \emptyset$.  
For every $i$ and $j$, $1 \leq i,j \leq 3$, on the vertices of the graph
$G_{i,j}$ we put the sets $S(w)$ with the vertices for which $f$ is equal
to $2$ receiving another color as follows:
to the vertices which belong to $X$ we add the color $i$, 
whereas to the vertices which belong to $Y$ we add the color $j$. 
The vertices $u$ and $v$ are both given the set $\{1,2,3\}$.
Let $f$ be a proper vertex-coloring of $H''$ assigning to each vertex
a color from its set. Denote $f(u)=i$ and $f(v)=j$, then $f$ restricted 
to $G_{i,j}$ is a proper vertex-coloring of $G$ assigning to each vertex
$w \in V$ a color in $S(w)$. $\Box$ 

\noindent
{\bf Proof of Theorem~\ref{t61}}\,
The proof is by induction on $k$. For $k=3$, the result follows from
lemma~\ref{l62}. Assuming that the result is true for $k$, $k \geq 3$, we
prove it is true for $k+1$. It is easy to see that 
{\bf BG $(k+1)$-CH} $\in \Pi_2^p$. We transform {\bf BG $k$-CH} to
{\bf BG $(k+1)$-CH}. Let $G=(V,E)$ be an instance of {\bf BG $k$-CH}.
We shall construct a bipartite graph $W$ such that $W$ is $(k+1)$-choosable
if and only if $G$ is $k$-choosable. 

Let $H$ be the disjoint union of the graphs 
$\{G_{i,j} : 1 \leq i,j \leq k+1\}$, where each $G_{i,j}$ is a copy 
of $G$. Let $(X,Y)$ be a bipartition of the bipartite graph $H$. 
The graph $W$ is obtained from $H$ by adding two new vertices 
$u$ and $v$, joining $u$ to every vertex of $X$, and joining $v$ to
every vertex of $Y$.

It is easy to see that if $G$ is $k$-choosable, then $W$ is $(k+1)$-choosable.
In a similar way to the proof of lemma~\ref{l62}, we can prove that if $W$ is
$(k+1)$-choosable, then $G$ is $k$-choosable. $\Box$

\section{The strong choice number}

Let $G=(V,E)$ be a graph, and let $V_1, \ldots ,V_r$ be pairwise disjoint
subsets of $V$. We denote by $[G,V_1, \ldots ,V_r]$ the graph obtained
from $G$ by appending the union of cliques induces by each  
$V_i$, $1 \leq i \leq r$.

Suppose that $G=(V,E)$ is a graph with maximum degree at most $1$. We claim
that $G$ is strongly $k$-choosable for every $k \geq 2$. To see that, 
let $V_1, \ldots ,V_r$ be pairwise disjoint subsets of $V$, each of size
at most $k$. The graph $[G,V_1, \ldots ,V_r]$ has maximum degree at   
most $k$, and therefore by corollary~\ref{c43} it is $k$-choosable.

\noindent
{\bf Proof of Theorem~\ref{t71}}\,
Let $G=(V,E)$ be a strongly $k$-colorable graph. Let $V_1, \ldots ,V_r$ be
pairwise disjoint subsets of $V$, each of size at most $k+1$. 
Without loss of generality, we can assume that $V_1, \ldots ,V_m$ are 
subsets of size exactly $k+1$, and $V_{m+1}, \ldots ,V_r$ are subsets
of size less than $k+1$. Let $H$ be the graph $[G,V_1, \ldots ,V_r]$. 
To complete the proof, it suffices to show that $H$ is $(k+1)$-colorable.  
For every $i$, $1 \leq i \leq m$, we define $W_i=V_i-\{c\}$ for an     
arbitrary element $c \in V_i$, whereas for every $j$, $m+1 \leq j \leq r$,
we define $W_i=V_i$. Since $[G,W_1, \ldots ,W_r]$ is $k$-colorable, there 
exists an independent set $S$ of $H$ which is composed of exactly one vertex
from each $V_i$, $1 \leq i \leq m$. For every $i$, $1 \leq i \leq m$, we
define $W'_i=V_i-S$, whereas for every $j$, $m+1 \leq j \leq r$, we 
define $W'_i=V_i$. Since $[G,W'_1, \ldots ,W'_r]$ is $k$-colorable, we can
obtain a proper $(k+1)$-vertex coloring of $H$ by using $k$ colors for
$V-S$ and another color for $S$. $\Box$

\begin{lemma}
\label{l71}
Suppose that $k,l \geq 1$. If ${\cal F}$ is a family of $k+l$ sets
of size $k+l$, then it is possible to partition ${\cal F}$ into a family 
${\cal F}_1$ of $k$ sets and a family ${\cal F}_2$ of $l$ sets, to 
choose for each set $S \in {\cal F}_1$ a subset $S' \subseteq S$ of size $k$,
and to choose for each set $T \in {\cal F}_2$ a subset $T' \subseteq T$ of
size $l$, so that $S' \cap T' = \emptyset$ for every $S \in {\cal F}_1$
and $T \in {\cal F}_2$.
\end{lemma}
{\bf Proof}\,
Suppose that ${\cal F} = \{C_1, \ldots , C_{k+l} \}$, and define 
$C = \cup_{i=1}^{k+l} C_i$. For every partition $\pi$ of $C$ into the
two subsets $A$ and $B$, we denote 
${\cal R}(\pi) = \{V \in {\cal F} : |V \cap A| > k \}$,
${\cal L}(\pi) = \{ V \in {\cal F} : |V \cap B| > l \}$,
and ${\cal M}(\pi)= \{ V \in {\cal F} : |V \cap A|=k~~and~~|V \cap B|=l \}$.  
We now start with the partition of $C$ into the two subsets $A=C$ and 
$B=\emptyset$, and start moving one element at a time from $A$ to $B$ 
until we obtain a partition $\pi_1$ of $C$ into the two subsets $A$ and $B$ 
and a partition $\pi_2$ into the two subsets $A'=A-\{c\}$ and $B'=B \cup \{c\}$,
such that $|{\cal R}(\pi_1)| > k$ and $|{\cal R}(\pi_2)| \leq k$.  
It is easy to see that 
${\cal L}(\pi_2) \subseteq {\cal L}(\pi_1) \cup {\cal M}(\pi_1)$,    
and therefore $|{\cal L}(\pi_2)| < l$. 
We now partition ${\cal M}(\pi_2)$ into two subsets ${\cal M}_1$ and 
${\cal M}_2$, such that ${\cal F}_1 = {\cal R}(\pi_2) \cup {\cal M}_1$
has size $k$ and ${\cal F}_2 = {\cal L}(\pi_2) \cup {\cal M}_2$
has size $l$. 
For every set $S \in {\cal F}_1$ we choose a subset 
$S' \subseteq S \cap A'$ of size $k$, whereas for every $T \in {\cal F}_2$  
we choose a subset $T' \subseteq T \cap B'$ of size $l$.
Since $A'$ and $B'$ are disjoint, we have that $S' \cap T' = \emptyset$
for every $S \in {\cal F}_1$ and $T \in {\cal F}_2$. $\Box$
\begin{lemma}
\label{l72}
Suppose that $k,m \geq 1$. If ${\cal F}$ is a family of $km$ sets
of size $km$, then it is possible to partition ${\cal F}$ into the  
$m$ subsets ${\cal F}_1, \ldots , {\cal F}_m$, each of
size $k$, and to choose for each set $S \in {\cal F}$ a subset 
$S' \subseteq S$ of size $k$, so that $S' \cap T' = \emptyset$ for
every $i \neq j$, $S \in {\cal F}_i$ and $T \in {\cal F}_j$.
\end{lemma}
{\bf Proof}\,
By induction on $m$. For $m=1$ the result is trivial. Assuming that
the result is true for $m$, $m \geq 1$, we prove it is true for $m+1$.
Let ${\cal F}$ be a family of $k(m+1)$ sets of size $k(m+1)$. 
By lemma~\ref{l71}, it is possible to partition ${\cal F}$ into
a family ${\cal F}_1$ of $k$ sets and a family ${\cal F}_2$ of $km$ sets,
to choose for each $S \in {\cal F}_1$ a subset $S' \subseteq S$ of size $k$,
and to choose for each set $T \in {\cal F}_2$ a subset $T' \subseteq T$
of size $km$, so that $S' \cap T' = \emptyset$ for every $S \in {\cal F}_1$
and $T \in {\cal F}_2$. The proof is completed by applying the induction
hypothesis on ${\cal F}_2$. $\Box$

\noindent
{\bf Proof of Theorem~\ref{t72}}\,
Let $G=(V,E)$ be a strongly $k$-choosable graph. Let $V_1, \ldots ,V_r$ be
pairwise disjoint subsets of $V$, each of size at most $km$. Let $H$ 
be the graph $[G,V_1, \ldots ,V_r]$. To complete the proof, it suffices to 
show that $H$ is $km$-choosable. For each $v \in V$, let $S(v)$ be a set of 
size $km$. By lemma~\ref{l72}, for every $i$, $1 \leq i \leq r$, is it
possible to partition $V_i$ into the $m$ subsets $V_{i,1}, \ldots ,V_{i,m}$,
each of size at most $k$, and to choose for each vertex $v \in V_i$
a subset $C(v) \subseteq S(v)$ of size $k$, so that 
$C(u) \cap C(v) = \emptyset$ for every $p \neq q$,
$u \in V_{i,p}$ and $v \in V_{i,q}$.
Since the graph $[G,V_{1,1}, \ldots ,V_{r,m}]$ is $k$-choosable, we can 
obtain a proper vertex-coloring of $H$ assigning to each vertex a color 
from its set. $\Box$ 

\noindent
{\bf Proof of Theorem~\ref{t73}}\,
Apply lemma~\ref{l72} as in proof of theorem~\ref{t72}. $\Box$

\noindent
{\bf Proof of Corollary~\ref{c71}}\,
It is proved in~\cite{FS} that if $G$ is a $4$-regular graph on $3n$ vertices
and $G$ has a decomposition into a Hamiltonian circuit and $n$ pairwise
vertex disjoint triangles, then $ch(G)=3$. The result follows from
theorem~\ref{t73}. $\Box$

\noindent
{\bf Proof of Theorem~\ref{t74}}\,
Since $s\chi(1)=2$, we can assume that $d>1$. Suppose first that $d$ is even,
and denote $d=2r$. Construct a graph $G$ with $12r-3$ vertices,
partitioned into $8$ classes, as follows. Let these classes
be $A,B_1,B_2,C_1,C_2,D_1,D_2,E$, where $|A|=|D_1|=|D_2|=2r$, $|B_1|=|B_2|=r$,
$|C_1|=|C_2|=r-1$, and $|E|=2r-1$. Each vertex in $A$ is joined by edges to
each member of $B_1$ and each member of $B_2$. Each member of $D_1$ is 
adjacent to each member of $D_2$. Consider the following partition of the
vertex set of $G$ into three classes of cardinality $4r-1$ each:
\begin{equation*}
V_1=B_1 \cup C_1 \cup D_1, V_2=B_2 \cup C_2 \cup D_2, V_3=A \cup E.
\end{equation*}

We claim that $H=[G,V_1,V_2,V_3]$ is not $(4r-1)$-colorable. In a proper
$(4r-1)$-vertex coloring of $H$, every color used for coloring the vertices 
of $A$ must appear on a vertex of $C_1 \cup D_1$ and on a vertex of 
$C_2 \cup D_2$. Since $|C_1 \cup C_2| < |A|$, there is a color used for 
coloring the vertices of $A$ which appears on both $D_1$ and $D_2$. 
But this is impossible as each vertex in $D_1$ is adjacent to 
each member of $D_2$. Thus $s\chi(G) > 4r-1$ and as the maximum degree 
in $G$ is $2r$, this shows that $s\chi(2r) \geq 4r$. 

Suppose next that $d$ is odd, and denote $d=2r+1$. Construct a graph $G$ 
with $12r+3$ vertices, partitioned into $8$ classes, as follows. Let these
classes be named as before, where $|A|=|D_1|=|D_2|=2r+1$, $|B_1|=r+1$,
$|C_1|=r-1$, $|B_2|=|C_2|=r$, and $|E|=2r$. In the same manner we can prove
that $[G,V_1,V_2,V_3]$ is not $(4r+1)$-colorable. Thus $s\chi(G) > 4r+1$ 
and as the maximum degree in $G$ is $2r+1$, this shows that
$s\chi(2r+1) \geq 4r+2$, completing the proof. $\Box$

\noindent
{\bf Acknowledgment}\,
We would like to thank Noga Alon for helpful discussions.

\end{document}